\documentclass{emulateapj}
\usepackage{graphicx,epsfig}

\usepackage{color}
\slugcomment{Submitted to Astrophysical Journal}

\shorttitle{Gas Feedback on Bar Evolution}
\shortauthors{Berentzen et al.}

\def\gtorder{\mathrel{\raise.3ex\hbox{$>$}\mkern-14mu
    \lower0.6ex\hbox{$\sim$}}}
\def\ltorder{\mathrel{\raise.3ex\hbox{$<$}\mkern-14mu
    \lower0.6ex\hbox{$\sim$}}}

\begin{document}

\title{GAS FEEDBACK ON STELLAR BAR EVOLUTION}

\author{Ingo Berentzen\altaffilmark{1}, Isaac Shlosman and Inma Martinez-Valpuesta\altaffilmark{2,3}}

\altaffiltext{1}{Present address: Astronomisches Rechen-Institut, Zentrum
f\"ur Astronomie, D-69120 Heidelberg, Germany}
\altaffiltext{2}{Gruber Foundation Fellow at OAMP, 13004, Marseille, France}
\altaffiltext{3}{Present address: Instituto de Astrofisica de Canarias, E-38200, 
La Laguna, Tenerife, Spain}
\affil{
Department of Physics and Astronomy,
University of Kentucky, Lexington, KY 40506-0055, USA \\ 
email: {\tt iberent@pa.uky.edu, shlosman@pa.uky.edu, martinez@pa.uky.edu}
}
\and
\author{Clayton H. Heller}
\affil{
Department of Physics, Georgia Southern University,
        Statesboro, GA 30460, USA \\
email: {\tt cheller@georgiasouthern.edu}
}

\begin{abstract}
We have analyzed evolution of live disk-halo systems in the presence
of various fractions of gas, $f_{\rm gas} \leq 8\%$ of the disk mass, for 5~Gyr. 
Specifically, we have addressed the issue of angular momentum ($J$) 
transfer from the gas to the stellar bar and its effect on the bar 
evolution. We find that the weakening of the bar over this time period, reported in the 
literature, is not related to the $J$-exchange with the gas, but is caused 
by the vertical buckling instability in the gas-poor disks and by a 
steep heating of a stellar velocity dispersion by the central mass 
concentration (CMC) in the gas-rich disks. On the other hand, the gas has a profound
effect on the onset of the buckling --- larger $f_{\rm gas}$ brings it forth
due to the more massive CMCs. The former process leads to 
the well-known formation of the boxy/peanut-shaped bulges, while the 
latter results in the formation of progressively more elliptical bulges, for
larger $f_{\rm gas}$. The subsequent (secular) 
evolution of the bar differs --- the gas-poor models exhibit a growing bar 
while gas-rich models show a declining bar whose vertical swelling is 
driven by a secular resonance heating. The border line between the 
gas-poor and -rich models lies at $f_{\rm gas}\sim 3\%$ in our models, 
but is model-dependent and  will be affected by additional processes, like star 
formation and feedback from stellar evolution. The overall effect of the gas 
on the dynamical and secular evolution of the bar is not in a direct $J$ transfer
to the stars, but in the loss of $J$ by the gas and its influx to the center
that increases the CMC. The more massive CMC damps the vertical buckling instability in 
the bar and depopulates orbits responsible for the appearance of 
boxy/peanut-shaped bulges. 
The combined action of resonant and non-resonant processes in gas-poor
and gas-rich disks leads to a converging evolution in the vertical extent of the bar
and its stellar dispersion velocities, and to a diverging evolution
in the bulge properties.
\end{abstract}


\section{Introduction}

Galactic bars break the axial symmetry of rotating disks in the most profound way
because they are sufficiently massive --- thus being of a paramount importance
to the short and long term galaxy evolution. However, various processes
associated with bars are still poorly understood. Barred galaxies consist of 
stellar disks embedded in the dark matter (DM) halos with an admixture of gas.
This gas is deprived of the rotational support by the bar and is channeled towards 
the central regions of few $\times 100~{\rm pc}-1$~kpc, accumulating there in the 
form of the Central Mass Concentration (CMC), that includes gas, stars and a certain
amount of the DM, and typically harbors the central
supermassive black hole (SBH). Our main goal is to analyze the 
changes in the stellar bar as a result of this process --- in other words to study 
the gas feedback on the bar evolution. 

In self-gravitating systems with a dynamically significant rotation, any substantial
departure from axial symmetry is destined to speed up the evolution, leading
to mass and angular momentum redistribution in the system.  
While the role of the angular momentum in this process has been emphasized 
already by Lynden-Bell \& Kalnajs (1972), the efficiency of this
transport and dependence on various parameters is still under investigation (e.g., 
Athanassoula 2003). Tremaine \& Weinberg (1984) and Weinberg (1985) have estimated 
that the bar should lose most of its momentum to the surrounding DM halo in a few 
rotations ($\sim 1$~Gyr) --- this was not supported by subsequent numerical 
simulations (Sellwood 2006), but the overall trend, that the angular momentum 
flows from the inner,
bar-unstable disk to the outer disk and to the halo, has been confirmed.            

Our view on the role of a DM halo in disk galaxy evolution has changed
dramatically over the last few years --- from the original claim that it
damps the bar instability (e.g., Ostriker \& Peebles
1973), to recent results that more massive halos grow larger
bars (e.g., Athanassoula \& Misiriotis 2002). It is a common wisdom today
that bars on all scales redistribute mass and angular momentum in the main 
body of the galaxy and alter the radial profiles of gas, stars
and DM densities. In the long run, the DM halos
serve as a sink for the angular momentum from the disk and this
process is mediated by the bars.

Disk galaxies possess various amounts of a cold gas, typically $< 10\%$
of the disk mass,
and probably have been more gas-rich in the past.
However, the ability of the gas to influence the galactic dynamics
of the parent object extends well beyond its mass fraction.
The gas dissipates and therefore can form bound massive
accumulations even in excess of $10^7~{\rm M_\odot}$. In this
sense the gas can be more clumpy than a collisionless
matter, either stellar or DM. This leads to a number
of dynamical consequences, more importantly to a dynamical
friction and to scattering and randomizing of stellar and
DM particle orbits (Shlosman \& Noguchi 1993). The angular momentum
redistribution in numerical simulations with gas has been
examined by Berentzen al. (1998) and Berentzen et al. (2004).

Furthermore, the gas bears similar amounts of a {\it
specific} angular momentum with stars in the cold disk 
prior to the bar instability.
Because of its viscosity, the gas responds to a
bar-like perturbation with a phase shift, compared to the
stellar response. This leads to streamline intersections
and to shocks downstreams from the bar major axis. Resulting 
leading dust lanes delineate the
underlying shocks and associated density enhancements (e.g.,
Athanassoula 1992). The 
gravitational torques from the bar extract the angular
momentum from the gas and transfer it to the underlying
stellar (and DM) component which lags behind the gas.
These torques depend on the shift
in the position angle between the gas and stellar distributions 
in the bar,

$$ {\rm Torque} = - \left[\int_0^{2\pi}\int_0^\infty \Sigma^{\rm a} (r,\phi) 
      {{\partial \Phi_{\rm gas}^{\rm a} (r,\phi)} \over {\partial \phi}} 
          \right] r dr d\phi ,    \eqno(1)  $$

\noindent where only the {\it asymmetric} part of the gas gravitational 
potential $\Phi_{\rm gas}^{\rm a}$ which acts on the segment $dr$ of the 
{\it asymmetric} stellar density distribution $\Sigma^{\rm a}(r,\phi)$ 
in the cylindrical system of coordinates $r, \phi$, will make a 
non-zero contribution. 

As a result, the gas falls towards the central kpc where
the bar potential is again more axisymmetric and the resulting
shock focusing injects the gas onto the weakly elliptical orbits {\it 
in situ}, forming nuclear rings. These rings are ubiquitous in barred
galaxies (Buta \& Combes 1996; Knapen 2005), but their subsequent evolution can
differ (Knapen et al. 1995; Heller \& Shlosman 1996; Heller, Shlosman \& 
Englmaier 2001; Englmaier \& Shlosman 2004). The ultimate fate
of the inflowing gas is debatable, but there is a strong
theoretical and observational evidence that some of this gas
can reach deep inside the central region and fuel the nuclear star
formation and the accretion processes onto the central SBH as a
result of gravitational instabilities in the gas itself
(e.g., Shlosman, Frank \& Begelman 1989; Shlosman, Begelman \& Frank
1990; Ishizuki et al. 1990; Kenney et al. 1992; Forbes et al. 1994; 
Knapen et al. 1995; 
Maiolino et al. 2000; Jogee et al. 2002; Shlosman 2005; Jogee 2006), and 
contribute to the formation of the BH itself (Begelman, Volonteri \& Rees 
2006; Heller, Shlosman \& Athanassoula 2007).

Whether the inflowing gas fuels the central SBH, or contributes
to the buildup of stellar bulges (Kormendy \& Kennicutt 2004 and refs. therein), 
the growing CMC in barred galaxies 
has been reported to have a strong effect on the stellar bar. 
Hasan \& Norman (1990), Friedli (1994), Norman, Sellwood \& Hasan (1996), 
Berentzen et al. (1998), Shen \& Sellwood (2004), and others have observed 
that the bar dissolves only when massive and compact CMCs form --- 
a process that is affected by the bar structure and its host DM halo 
(Athanassoula, Lambert \& Dehnen 2005). Such extreme CMCs can be represented 
only by the SBHs. However, the required SBH mass of a few percent of the disk 
mass is more than a factor of ten larger than the SBH masses found in disk 
galaxies (e.g., Ferrarese \& Ford 2005). Hence the stellar bars are much more 
resilient than previously envisioned.

Furthermore, Bournaud, Combes \& Semelin (2005) have argued, based on
numerical simulations of 7.25\% gas-rich disks embedded in {\it rigid} 
halos, that the gas is able to weaken the stellar bars dramatically, even
{\it before} the CMC is in place --- the reason for this is
the transfer of angular momentum from the gas to the bar.
The combination of angular momentum transfer to the stars and the
subsequent buildup of the CMC destroy the bar over the timescale as 
short as 1.4~Gyr --- claimed by Bournaud et al. 

Here we analyze the bar evolution for various gas fractions
in the disk and allow for the growth of the CMC and SBH.
Unlike Bournaud et al. (2005), we use a disk immersed in the live
DM halo. In addition to the standard model, we supplement our 
analysis with a number of test models. We limit the discussion
to 5~Gyr of bar evolution in order to make a direct comparison
with Bournaud et al. This paper is structured as following. Section~2 
describes our numerical
tools and the initial conditions used in this work.
Section~3 provides the results and Section~4
presents a number of test models. We conclude with the discussion
in Section~5.

\section{Numerical method and Initial Conditions}

We use the updated hybrid $N$-body and Smooth Particle hydrodynamics (SPH) code
of Heller \& Shlosman (1994). The version FTM-4.4 of the code uses the FalcON
force solver of Dehnen (2002) --- a tree code with mutual cell-cell interactions
and complexity $O(N)$. It conserves the momentum exactly and is
about ten times faster than the optimally-coded Barnes \& Hut (1986) tree code.
We use a constant gravitational softening of 160~pc for
the gas and for the collisionless particles. 
The amount of DM particles is $N_{\rm DM}=1.2\times 10^5$, stellar particles  
$N_*=3.6\times 10^5$, and gas $N_{\rm gas}=4\times 10^4$. The DM particle
mass is $1.1\times 10^6~{\rm M_\odot}$, the stellar particle is 
$1.5\times 10^5~{\rm M_\odot}$, and the gas particle is $1.2\times 10^5~{\rm M_\odot}$
for the gas fraction $f_{\rm gas} = 8\%$ of the disk mass. Models with a much larger 
number of the disk and DM particles, making DM/stellar mass ratio per particle $\sim 1$,
did not change the evolution over the run times of our models. The energy and angular
momentum concervation in the pure collisionless models is better than 0.15\% and 0.02\%,
correspondingly.
 
\begin{deluxetable*}{lccccccccc}[ht!!!!!!!!!!!!!!]
\tablecaption{\bf G0: Standard Model Parameters}
\tablehead{
Parameters         & Stellar Disk & DM Halo & SBH    & Notes \nl }
\startdata
%
Radial scalelength  &    2.85 kpc  &  ---   &  ---   & \nl
Vertical scalelength&    0.2 kpc   &  ---   &  ---   & \nl
Total mass &$0.58\times 10^{11}~{\rm M_\odot}$&$1.33\times 10^{11}~{\rm M_\odot}$&0& \nl
Mass ($<10$~kpc)    &$0.5\times 10^{11}~{\rm M_\odot}$&$0.5\times 10^{11}~{\rm M_\odot}$& 0 & \nl
Gas fraction ($f_{\rm gas}$)&      0       &    0   &  ---   & \nl
\enddata
\label{table:models}
\end{deluxetable*}

\begin{deluxetable*}{lcccclccl}[ht!!!!!!!!!!!!!!!!!!!!!!!!!!]
\tablecaption{\bf List of Models}
\tablehead{
Model & $f_{\rm gas}$ (\%) & SBH properties & Q & Main Figs. & Notes \nl }
\startdata

{\bf G0}      &  0  & ---     & 1.5 & 1a,b;& Standard Model: DM + stellar disk \\
\hline
{\bf G05}     & 0.5 & ---     & 1.5 & 1a,b;2 & Standard Model with 0.5\% gas\\
{\bf G2}      &  2  & ---     & 1.5 &      & as G05 with 2\% gas\\
{\bf G4}      &  4  & ---     & 1.5 &      & as G05 with 4\% gas\\
{\bf G6}      &  6  & ---     & 1.5 &      & as G05 with 6\% gas\\
{\bf G8}      &  8  & ---     & 1.5 &      & as G05 with 8\% gas\\
\hline
{\bf G05BH}   & 0.5 & growing & 1.5 & 1a,b;3 & as G05 but with a growing SBH\\
{\bf G2BH}    &  2  & growing & 1.5 &      & as G2 but with a growing SBH\\
{\bf G4BH}    &  4  & growing & 1.5 &      & as G4 but with a growing SBH\\
{\bf G6BH}    &  6  & growing & 1.5 &      & as G6 but with a growing SBH\\
{\bf G8BH}    &  8  & growing & 1.5 &      & as G8 but with a growing SBH\\
\hline
{\bf G8-25T}  & 8   & growing & 1.5 & 4    & 25\% gas torques removed \\
{\bf G8-50T}  & 8   & growing & 1.5 &      & 50\% gas torques removed \\
{\bf G8-75T}  & 8   & growing & 1.5 &      & 75\% gas torques removed \\
{\bf G8-100T} & 8   & growing & 1.5 &      & 100\% gas torques removed \\
\hline
{\bf G0GT}    &     & ---     & 1.5 & 5   & no gas, adding 'quasi-exact' gas force field in \\
              &     &         &     &     & \ \ \ \ form of an external potential from G8BH\\ 
\hline
{\bf G0BH}    & 0   & growing & 1.5 & 6   & using G8BH model for the SBH growth \\
{\bf G0BH250} & 0   & growing & 1.5 & 6   & using G8BH model for the SBH growth and gas \\ 
              &     &         &     &     & \ \ \ \ inside $r_0=250$~pc added to the SBH \\
{\bf G0BH80}& 0   & growing & 1.5 & 7      & using G8BH model: SBH mass incl. gas within 250 pc. \\
              &     &         &   &        & \ \ \ \ The only model that has $\epsilon_\bullet = 80$~pc  \\
{\bf G0BH750} & 0   & growing & 1.5 & 7    & using G8BH for the SBH and gas (Plummer sphere \\ 
              &     &         &     &      & \ \ \ \ within $r_0=750$~pc) growth \\
\hline
{\bf G0-Q18}  & 0   & ---     & 1.8 & 8   & as G0 but with $Q=1.8$ \\ 
{\bf G8BH-Q18}& 8   & growing & 1.8 & 8   & as G8BH but with $Q=1.8$ \\
\enddata
\label{table:models}
\tablecomments{Columns: (1) model type (see text); (2) gas fraction (\%) of the disk mass; 
(3) ``growing''
--- growing the BH from a seed value of $10^5~{\rm M_\odot}$; (4) initial value of
$Q$ parameter; (5) main figure(s) introducing this model; (6) comments}
\end{deluxetable*}

For models with SBHs, the central SBH evolution is given by a single
stellar particle with initially a small `seed' mass of $10^5~{\rm M_\odot}$. It is
nailed to the position of the center-of-mass (CoM) of the system at $\tau=0$ and
has a fixed 
gravitational softening $\epsilon_\bullet = 160$~pc (a Plummer sphere with a 
characteristic $r_0=160$~pc), except in one model (G0BH80, see Table~2) where
$\epsilon_\bullet = 80$~pc. The CoMs of the disk and the halo stay within $\sim 50$~pc
from the SBH, damping each others motions. We find no differences in the behavior
of the CoMs in the models with and without the gas.
The accretion radius of the SBH is taken as $R_{\rm acc}=40$~pc.
Particles which are found within this radius {\it and} which are bound to the SBH,
are extracted from the simulations and their mass is added to the SBH.
The requirement for a small timestep near the CMC and the central SBH are 
satisfied by the hierarchical timesteps in the FTM code. We typically use
between nine to twelve time bins, each differing by a factor of 2.

The initial conditions (see Table~1) have been obtained iteratively from Fall \&
Efstathiou (1980) to assemble disk and halo particles in a virial equilibrium (see Heller
\& Shlosman 1994 for more details). 
The halo has been relaxed in the frozen disk potential for $\Delta\tau \sim 2.4$~Gyr. 
The disk is axisymmetric with the Toomre's parameter $Q=1.5$ for the stars only, and the 
halo-to-disk mass ratio is $\sim 1$ within $R=10$~kpc, at the start of the 
simulations. Other values of $Q$ are used as well, and the asymmetric drift is 
accounted for. When 
the gas is added to the disk, the total disk mass is kept unchanged. We use
the isothermal equation of state (EOS) with the temperature of $T=10^4$~K but have run
also models with the adiabatic EOS --- no differences have been detected. The gas has
the stellar radial distribution initially, vertically it is in a hydrostatic equilibrium. 
The pure 
collisionless model parameters are listed in Table~1 and the list of all the
models with gas and additional test models  are listed in Table~2. Overall, our initial 
conditions are
identical to those used by Berentzen et al. (1998), but the numerical resolution
in the current models is much superior. This allows us to compare the models
and be guided by Berentzen et al. nonlinear orbit analysis and their usage of
the Poincare sections in dissecting the models.

\section{Results: Varying the Gas Fraction}

\begin{figure*}
\begin{center}
\includegraphics[angle=-90, width=0.7\columnwidth]{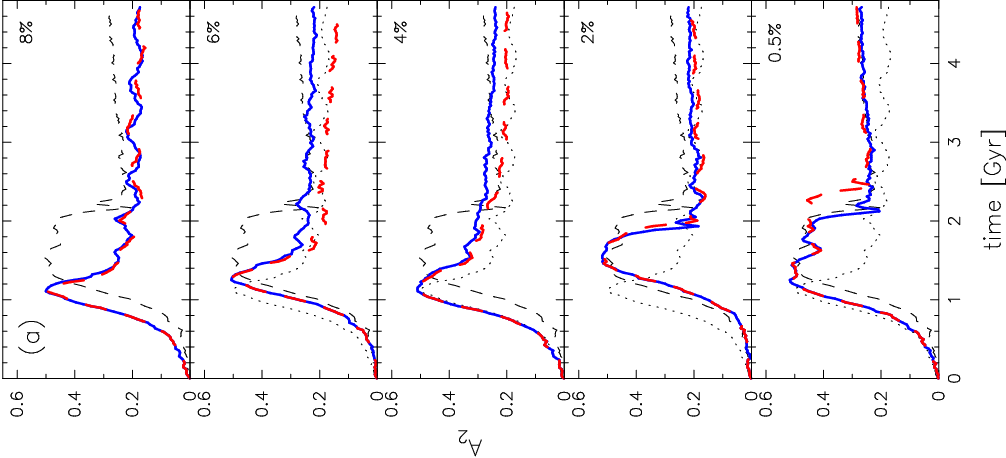}\hspace{2.cm}
\includegraphics[angle=-90, width=0.71\columnwidth]{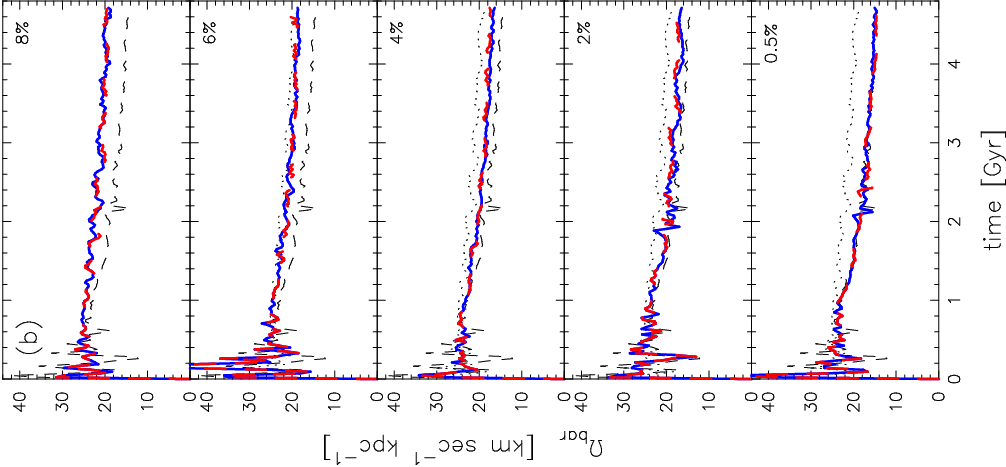}
\end{center}
\caption{Evolution of $(a)$ stellar bar amplitude $A_2$ and $(b)$ the bar pattern 
speed $\Omega_{\rm bar}$ (in km\,sec$^{-1}$\,kpc$^{-1}$) given by the $m=2$ mode in 
models (from the bottom to the top) G05--G8 (blue lines), with various 
$f_{\rm gas}$, superposed on the 
G05BH--G8BH (red dashed lines) with a growing SBH, and on the collisionless model G0 
(black dashed line) over the first 5~Gyr. For a comparison, we also display G8 model
(dotted black lines).}
\end{figure*}
\begin{figure}
\begin{center}
\includegraphics[angle=-90,width=0.8\columnwidth]{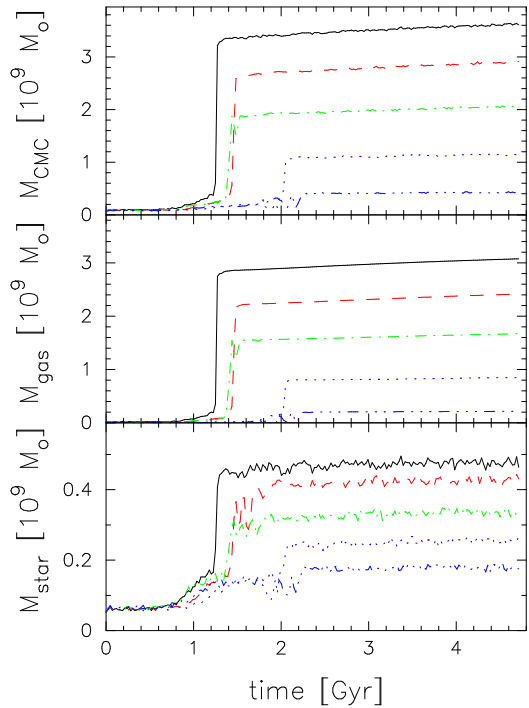}
\end{center}
\caption{Growth of the CMC in G05 --- G8 models within the central 250~pc. Top:
total (CMC) mass within this radius; Middle: gas mass; Bottom: stellar mass. 
The colors are: G05 (blue dashed-dot-dot-dotted), G2 (blue dotted), G4 (green-dotted), G6 
(red-dashed) and G8 (solid black).
}
\end{figure}

Our benchmark model G0 is a pure collisionless (DM$+$stars) model which shows a 
rapidly growing stellar bar whose $m=2$ normalized amplitude $A_2$ of the stellar 
component saturates at $\tau\sim 1.2$~Gyr. We define $A_2$ within a cylindrical
region of $0 \leq r \leq 7.5$~kpc and $|z|\leq 1$~kpc, so it encompasses the 
modeled stellar bars at nearly all times, and normalize it by the amplitude of the $m=0$
term. After the extended plateau of about 0.9~Gyr, this amplitude drops sharply to 
about $A_2\sim 0.2$. The following evolution of the bar is that of a gradual 
strengthening (Fig.~1a). 

Models with $f_{\rm gas} = 0.5\%-8\%$ gas, G05--G8 (Fig.~1a), 
without the central BH, show a 
similar rise and drop as G0 in the bar strength. The main differences appear to be the
existence of an extended {\it plateau} before the drop in $A_2$, which gradually 
disappears in models with larger gas fraction, for $\gtorder 4\%$, and  
the maximum in $A_2$ which is reached slightly earlier for these models. 
The behavior of $A_2$
near its maximum changes from G05 to G8 gradually --- in the gas poor models
the bar evolution converges to that in G0. Following the large drop, $A_2$ changes
little over the time of the simulations: G0--G2 show a slight increase while G4--G8 
models show a slight decrease. Here we focus on the sharp drops in $A_2$ that are 
visible in all models. 

The bar pattern speeds, $\Omega_{\rm bar}$, are shown in Fig.~1b.  
Models with larger $f_{\rm gas}$ slow down more gradually than the gas-poor 
models, by about 30\%.  There is also a substantial 
difference between the behavior of the corotation (CR) radius in gas-rich and gas-poor 
models. In G0, the CR increases from $\sim 9$~kpc to 14~kpc, while that of G8 model 
stays flat initially and then increases negligibly to 10~kpc over the simulation time 
of 5~Gyr.

The CMCs which correspond to a total mass accumulation within the central 250~pc, as well
as their gaseous and stellar components, for models 
G05--G8, have been displayed in Fig.~2. The pure stellar model G0 does not grow a 
visible CMC, while other models grow it during 0.2--1.5~Gyr, with longest timescale
corresponding to the most gas-poor models. The growth period can be roughly divided into
the initial shallow growth --- this is more prolonged for models with lower $f_{\rm gas}$,
and the second, avalanche-type growth, that is $\sim 0.15$~Gyr for all models. The peak 
growth rate of the CMC is attained around the peak of the bar strength.

\subsection{Results: Adding the Central BH}

In models G05BH--G8BH, the bar-triggered radial gas inflow leads to the gas accumulation 
in the central region as well as in the fueling of the central BH surrounded by gas. 
Figs.~2 and 3 demonstrate both the growth of the BH mass, $M_\bullet (\tau)$, and that of the
CMC within the central 250~pc, $M_{\rm CMC250}(\tau)$, which includes the BH, gas, 
stars and DM, as a function of time. Subsequently to the bar decay from
its maximal strength, the growth of $M_\bullet (\tau)$ and of $M_{\rm CMC250}(\tau)$ 
saturate within $\ltorder 0.2$~Gyr, i.e., almost instantly.  The subsequent evolution 
of the CMC and the BH is very mild. The final $M_\bullet$ and $M_{\rm CMC250}$ scale 
linearly with the initial gas fraction in the disk. 
A fraction $\sim 65\%-75\%$ of the gas ends up in the CMC in all 
models with the BH --- for smaller $f_{\rm gas}$ there is fractionally more gas in the end
(Fig.~3). Overall, we do not find a substantial difference between the CMCs 
in models with and without the central BH. However, models with larger $f_{\rm gas}$
grow more massive BHs and CMCs affecting the subsequent bar evolution, both its
vertical and planar structures, as we discuss in Section~5.

\begin{figure}
\begin{center}
\includegraphics[angle=-90,width=0.8\columnwidth]{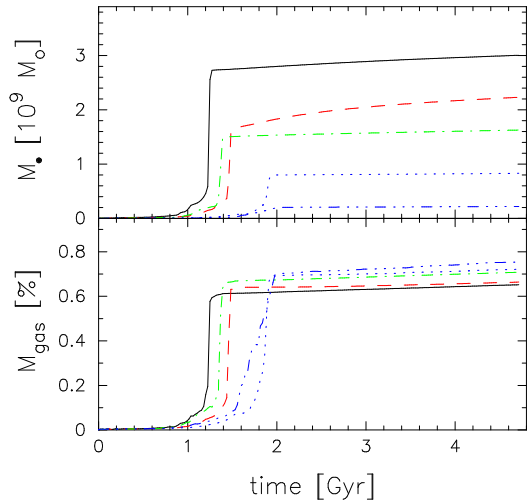}
\end{center}
\caption{Growth of the central BH (upper frame) in G05BH --- G8BH models,
and evolution of the gas fraction of the total CMC mass within the central 250~pc (lower 
frame). The mass of the BH is counted as contributing to the gas mass. The colors are: 
G05BH (blue dotted), G2BH (blue dashed-dot-dot-dotted), G4BH (green dash-dotted), G6BH (red-dashed) 
and G8BH (solid black).
}
\end{figure}

\section{Testing the Models}

Models with an increasing gas fraction, both with and without the BH, exhibit 
 continuity of bar properties.
All of them develop bars of nearly identical strength when measured by the $A_2$ amplitude.
Even more spectacular is the subsequent decrease in the bar strength as shown in
Fig.~1. While the shape of $A_2(\tau)$ differs among the models, the amount of the 
post-maximum drop is the same. Because the models differ in the gas fraction, $f_{\rm gas}$,
the evolution of the CMC and of the central
BH will differ as well. Comparison of Figs.~1--3 shows that the presence of the SBH has only 
a minimal influence 
on the bar strength, its pattern speed, and on the CMC. We, therefore,
consider the models with the BH as representative and base our discussion on them.
Difference between the models is mentioned only when it is substantial.

\subsection{Removing gas gravitational torques on stars}

The stellar and gas fluids differ in their intrinsic physical properties, 
specifically in viscosity, which generates the time lag in the gas response
to any perturbation. This time lag is the source of gas-stars 
mutual gravitational torques within the bar (Eq.~1) --- with
no viscosity the contribution of this integral is zero. The
angular momentum within the bar flows from the gas to the 
stars in the disk and to a certain degree to the DM halo. 
Removing these torques and comparing the models will allow for a direct testing 
of Bournaud et al. (2005) claim that they affect the stellar bar evolution.  

Because we aim at understanding the effect of the gas on the bar evolution, specifically
through angular momentum redistribution in the system, we use the bar strength measured
by $A_2$ and follow the balance of the angular momentum in the basic morphological
components --- the disk, bar and the DM halo. We first ask the question, to what degree 
the gas is responsible for the bar weakening as shown in the Fig.~1a frames. We test 
the possibility that the large drop in the
bar strength shown in various models of Fig.~1a between $\tau\sim 1-2.5$~Gyr results
from the input of the angular momentum coming from the gas gravitational torques 
as proposed by Bournaud et al. (2005).

\begin{figure}
\begin{center}
\includegraphics[angle=-90,width=0.9\columnwidth]{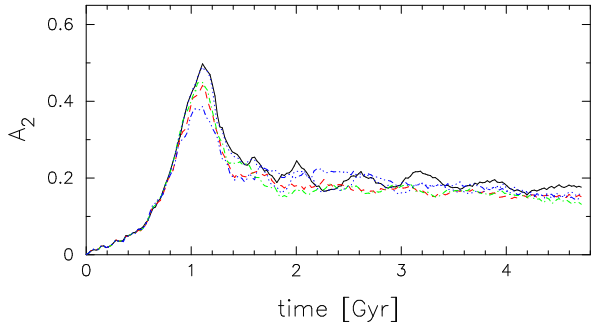}
\end{center}
\caption{Evolution of G8 model with gradually subtracted torques from the gas onto 
the stars. Shown are models with 25\% torques subtracted (black dotted), 50\% subtracted
(green dash-dotted), 75\% subtracted (red dashed), 100\% subtracted (blue dash-dot-dot-dotted). 
For the comparison we also display G8 (black solid).
  }
\end{figure}

The fact that G0, the pure stellar model, shows the same qualitative behavior as
other models and the degree of weakening is even quantitatively similar, rises 
serious doubts that the gas is in any way responsible for the bar downsizing. 
Nevertheless, we perform the first test of removing a fixed fraction of the 
gravitational torques, by removing the tangential components of the gravitational 
forces, applied by the gas on the stars and the DM. Fig.~4 exhibits four models based 
on G8BH where the torques have been reduced by 25\%, 50\%, 75\% and 100\%, i.e., 
models G8-25T--100T. It shows that there is no substantial difference between the
G8BH and G8-T models. While the $A_2$ peak around $1.1$~Gyr lowers slightly, when
the torques are removed, we consider this to be a numerical rather than physical
effect. For example, differences of this level are even expected for the same model
that uses a different random `seed' in the initial conditions. The subsequent
evolution of models in Fig.~4 shows a clear convergence trend.
While indeed the stellar bar receives less angular momentum from the
gas, one is forced to conclude that the 
gas torques have no effect on the observed drop in the bar strength in our models. The 
question of course is what is the origin of this drop. We address this issue in Section~5 
after exploring various venues via test models.  

\subsection{Gas Substitutes}

In tandem with tests shown in Fig.~4, we advanced the G0 model with 
artificially added gravitational torques from the gas in G8BH model. For this purpose,
we calculate the torques from the gaseous component on a cartesian grid for the (total) 
dimension of 50~kpc $\times$ 50~kpc $\times$ 4~kpc. The (constant) grid spacing is
250~pc along the $x$- and $y$-axes and 2~kpc along $z$.  
In order to have a smooth force field, we take time average over several
frames. We bring up the force field quasi-adiabatically between $\tau\sim 0.94-1.22$~Gyr (Fig.~5), or over 20 dynamical times,
reaching its full strength about the time when the bar amplitude also reaches
its maximum. To apply the torques on the stellar and DM components, we use
a 2D spline interpolation.

The resulting evolution (model G0GT) follows closely that of the G8BH model, with
the exception of the $A_2$ drop time, i.e., the extent of the associated plateau. 
Both tests performed in this section agree with our previous conclusion that, for models 
with $f_{\rm gas}\ltorder 8\%$, the gravitational torques from the gas do not alter substantially the drop in the $A_2$ observed in all models, but have a profound effect 
on the extent of the plateau which precedes the drop. They also serve as an independent 
verification that our modeling of effects of the gas component are 
sufficiently reasonable and do not alter the model evolution in some 
unexpected way.

\begin{figure}
\begin{center}
\includegraphics[angle=-90,width=0.9\columnwidth]{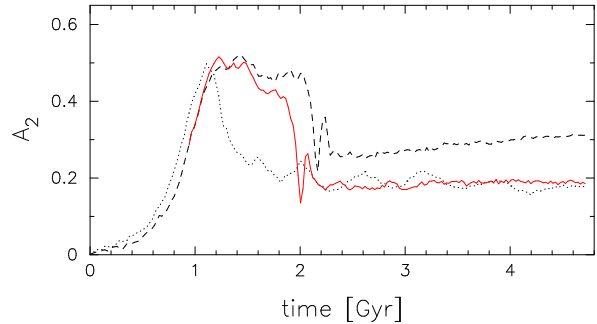}
\end{center}
\caption{Evolution of $A_2$ in the G0GT model (solid red): constructed from the G0 
model under the gravitational torques from the gas onto the stars in G8BH. For the 
comparison we also display G8BH (black dotted) and G0 (black dashed).}
\end{figure}

\subsection{Artificial growth of the BH and the CMC}

\begin{figure}
\begin{center}
\includegraphics[angle=-90,width=0.9\columnwidth]{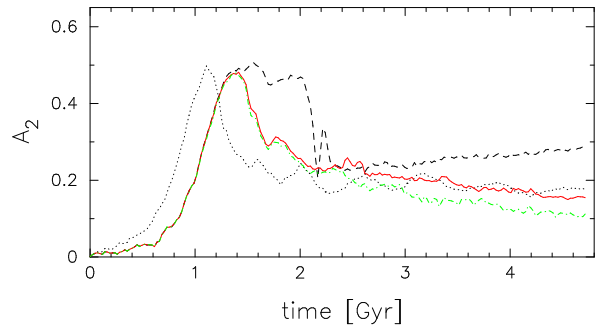}
\end{center}
\caption{Evolution of $A_2$ in (a) G0BH model with an artificially grown BH, whose
history is taken from G8BH (red line), and (b) G0BH250 model with an artificially 
grown BH and the gas within the central 250~pc (added to the BH mass) (green dash-dotted).
G0 and G8BH curves were added for a comparison.}
\end{figure}
\begin{figure}
\begin{center}
\includegraphics[angle=-90,width=0.9\columnwidth]{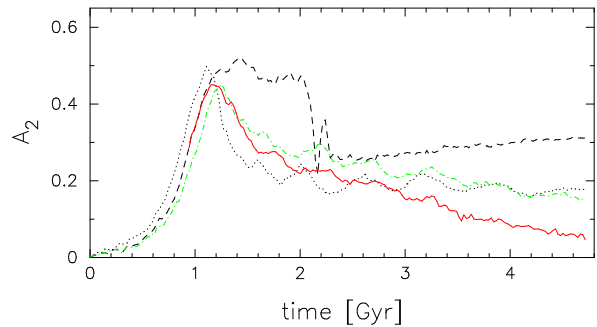}
\end{center}
\caption{Evolution of $A_2$ in (a) G0BH80 model with an artificially grown BH with a
smaller gravitational softening of 80~pc, whose
history is taken from G8BH and is a combined mass of the SBH and the gas within the
central 250~pc (red line), and (b) G0BH750 model with an artificially grown BH (regular
softening of 160~pc)
and the gas within the central $r_0=750$~pc which is modeled as a Plummer sphere
with $r_0$ (green dash-dotted). G0 and G8BH curves were added for a comparison.}
\end{figure}

The stellar bar is expected to evolve with respect to the buildup of the central 
BH and the nearby gas accumulation. If these grow on a secular timescale, i.e.,
adiabatically, this gas will drag in additional stars and DM. The evolution of 
the mass within the central 250 pc and the central SBH are shown in 
Figs.~2 and 3 for models with various $f_{\rm gas}$.

To further isolate the consequences of a gas influx towards the central regions, we 
have used the pure stellar model G0 and imposed the BH and gas accumulation histories 
taken from a gas-rich model. The CMC is defined as earlier and consists of the gas 
accumulation within the central 250~pc (or 750~pc), stars and DM there. Fig.~6 shows 
two such models: $(a)$ an artificially growing BH taken 
from the G8BH (model G0BH); $(b)$ an additional model of a growing BH from G8BH when 
the gas within the central 250~pc is added to the BH mass model (G0BH250). In both 
cases the gravitational softening of the BH is 160~pc --- the typical softening
in our models. The model G0BH closely follows the corresponding G8BH, although
no gravitational torques from the gas are present here. The $A_2$ curve has switched
gradually from that of G0 to G8BH. The corresponding model G0BH250 with more
massive SBH (as the gas within the central 250~pc has been added to the BH) falls
below the G0BH curve, as expected --- this confirms that more massive SBHs, albeit 
not found in disk galaxies by a large margin, can in fact dissolve the stellar 
bars, as discussed in the literature.  

Fig.~7 displays $(a)$ the evolution a BH with an added gas mass from the central 250~pc
but the gravitational softening of the BH is decreased to 80~pc (model G0BH80), to 
test a more compact mass distribution; and $(b)$ a growing BH and a independently 
growing the gas accumulation within the central 750~pc (model G0BH750). This gas
is approximated by a second Plummer sphere with a characteristic radius of 750~pc. 
The G0BH80 bar decays gradually and its $A_2$ fall below that of G0BH250 ---
smaller gravitational softening for the BH creates a more compact CMC here which
starts to affect the bar secularly. The substantially more massive CMC in 
G0BH750 has a profound effect on the bar by dissolving it during $\sim 5$~Gyr, as
discussed in section~1.

\subsection{Stellar Bars in Hot Disks} 

Stellar bar instability is delayed and its amplitude is lowered in hotter disks,
i.e., disks with larger stellar dispersion velocities (e.g., Athanassoula \& Sellwood
1983). In order to test this effect on the abrupt weakening of the bar in our models,
we have increased the initial $Q$ parameter to 1.8 in G0 and G8BH models, hereafter
models G0-Q18 and G8BH-Q18. Fig.~8  displays the evolution of these models. The risetime
of the bar instability has increased as expected in both models, and the $A_2$ peaks
are delayed with respect to the original models. The maxima
of the $A_2$ amplitudes have been also lowered by $\sim 0.1$, and the amplitude of the 
drop in $A_2$ has diminished by this amount as well. The difference between the above 
models after the amplitude drop decreases substantially. The stellar bars survive and 
do not show any sign of decay over the simulation time of 5~Gyr.

\begin{figure}
\begin{center}
\includegraphics[angle=-90,width=0.9\columnwidth]{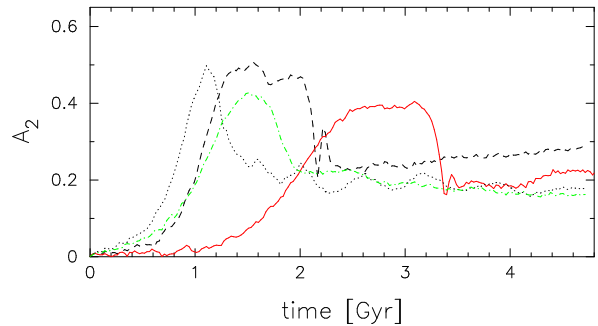}
\end{center}
\caption{Evolution of $A_2$ in G0-Q18 (red line) and G8BH-Q18 (green dash-dotted) 
models which are similar to G0 and G8BH but with $Q=1.8$. For the comparison we also 
display G8BH (black dotted) and G0 (black dashed). 
}
\end{figure}

\subsection{Angular Momentum Evolution} 

\begin{figure}
\begin{center}
\includegraphics[angle=-90,width=0.85\columnwidth]{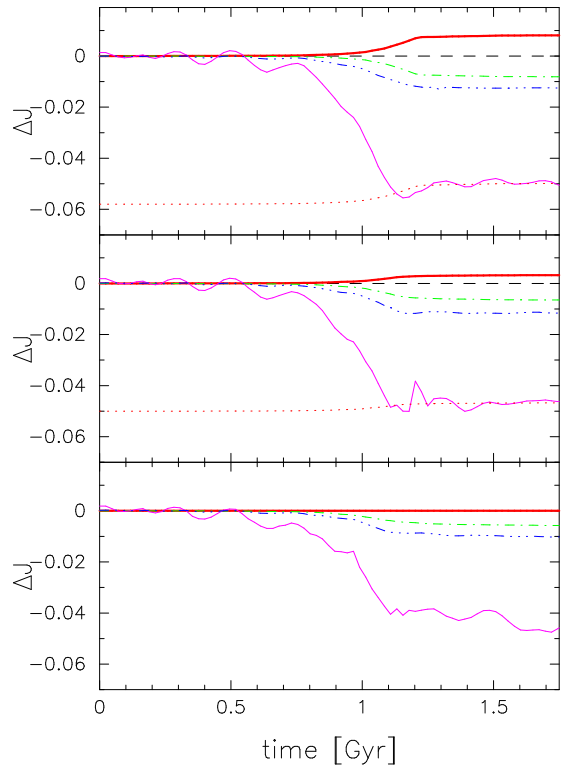}
\end{center}
\caption{Change in the angular momentum ($\Delta J$) within the CR radius in G8BH (top), 
G8-50T (center) and G8-100T (bottom). The solid red lines show $J$ flow from the gas
to the stars calculated from integration of corresponding gravitational torques.
The dash-dotted green lines provide the $J$ flow from the stars to the gas, calculated 
from the torques as well. The dash-dot-dot-dotted blue and solid pink lines exhibit the $J$ 
flow in the gaseous and stellar component, respectively, calculated directly from 
model evolution. The red dotted lines are identical to the red solid lines shifted to 
match the pink lines. The black dashed line is added to emphasize the $\Delta J = 0$ 
line.
}
\end{figure}

Next, we analyze the angular momentum ($J$) redistribution between the disk and the 
DM halo in the presence of the gas. The resonant interaction between the 
various morphological components will be addressed elsewhere. We have divided the 
disk/halo system into the (cylindrical) part within the CR and outside it. 
The overall $J$ of the DM halo increases sharply only after the bar has reached it 
maximal amplitude, i.e., after $\sim 1$~Gyr and is larger by a factor of $\sim 2$ 
for the gas-poor models than for G8. The inner and outer halos (with respect to the 
bar CR) follow the same trend, with the outer halo gaining more than the inner one.  

The angular momentum of the disk is decreasing steadily, more sharply after
$\sim 1$~Gyr, in all models. This decrease is a result of the
$J$ loss by the inner disk over this time period. Subsequently, this $J$ increases
very slightly in the gas-poor models and saturates in gas-rich models. The outer disk
behavior traces (anti-)symmetrically that of the inner disk. The gas in the inner
disk possesses small fraction of $J$ (i.e., in the stellar disk) even in gas-rich 
models evolved here and rapidly loses it
during the first Gyr, then stays flat for the remaining of the simulations.   
So, $J$ in the system flows from the inner disk to the outer disk and to 
the halo. The inner halo responds to this trend, largely because of the appearance 
of the `ghost' bar in the DM there (Athanassoula 2005; Berentzen \& Shlosman 2006). 

The role of gas in the overall balance of $J$ in the disk/halo system can be 
estimated from Fig.~9. Here we compare the direct input of $J$ into the stellar
disk via the gravitational torques from the gas with the total flow of $J$ in the
disk. The solid red lines show the input of $J$ by the gas to the stars by integrating
over the torques in models with gradually reduced torques, G8BH, G8-50T and G8-100T.
The solid pink lines describe the total $\Delta J$ in the stellar component. The 
red dotted line is identical to the red solid one but shifted downward for a direct 
comparison with the pink line. The $J$ input by the gas into the stars over 5~Gyr
of the run is $\sim 13\%$ of the stellar $J$ loss to the DM halo over this 
time. This relatively small input explains the near independence of the bar strength
on the gas torques. However, as we discuss in the next section, the gas accumulation
in the center resulting from the stellar bar has a profound effect on the evolution
of the vertical structure in the bar.

\section{Discussion: Bar Evolution and Gas}

The main goal of this work is to understand the effect of gas on stellar bar
evolution. For this purpose we have run a series of models for stellar disks embedded
in live DM halos, with various gas fractions ranging from 0\% to 8\% of the disk mass, 
with and without a 
growing central BH. In all of these models, the bar growth is barely affected by
the gas presence --- more gas-rich models display a marginally shorter rise time
of the bar instability. The maximal values of the bar amplitude, $A_2\sim 0.5$, attained
by the bar are very similar in all models. Past this maximum, the gas-poor models 
exhibit a plateau
in $A_2$ for about 1~Gyr, followed by a sharp drop in the amplitude to $\sim 0.2$.
This plateau gradually disappears with increasing $f_{\rm gas}$ and its extent
clearly anticorrelates with the CMC mass. Subsequently,
the models differ in their evolution, ranging from a slight
decline to a slight increase in $A_2$ --- this evolution clearly separates gas-poor 
(G0--G2; $f_{\rm gas} < 3\%$) from gas-rich (G4--G8; $f_{\rm gas} > 3\%$) models. 
In all models the stellar bar has survived
during the computation time of 5~Gyr, although appears to be substantially weakened. 
Models with growing BHs show no significant
difference in their evolution compared to models without the BH.  

\begin{figure}
\begin{center}
\includegraphics[angle=-90,width=0.90\columnwidth]{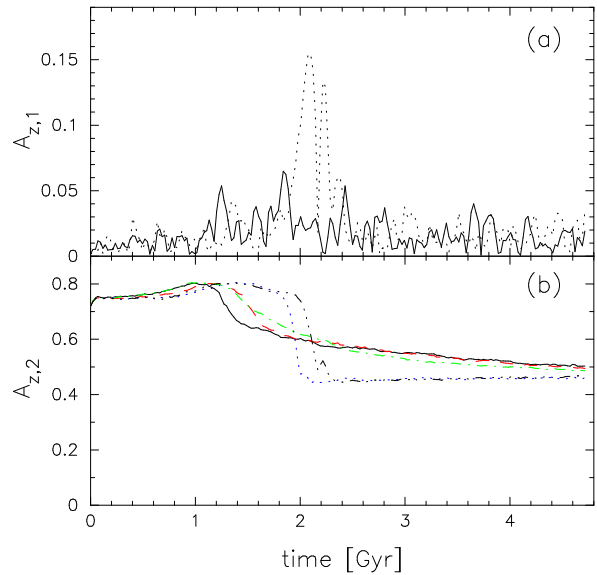}
\end{center}
\caption{$(a).$ The vertical asymmetry of the disk measured by $A_{\rm z,1}$ Fourier 
coefficient. Shown are G0 (dotted black line) to G8BH (solid black) models. 
$(b).$ The vertical `strength' of the stellar bar measured by $A_{\rm z,2}$ for
G8BH (solid black), G6BH (dashed red), G4BH (dash-dotted green), G2BH (dotted blue) 
and G0 (black dash-dot-dot-dotted) models.}
\end{figure}
\begin{figure}
\begin{center}
\includegraphics[angle=-90,width=0.85\columnwidth]{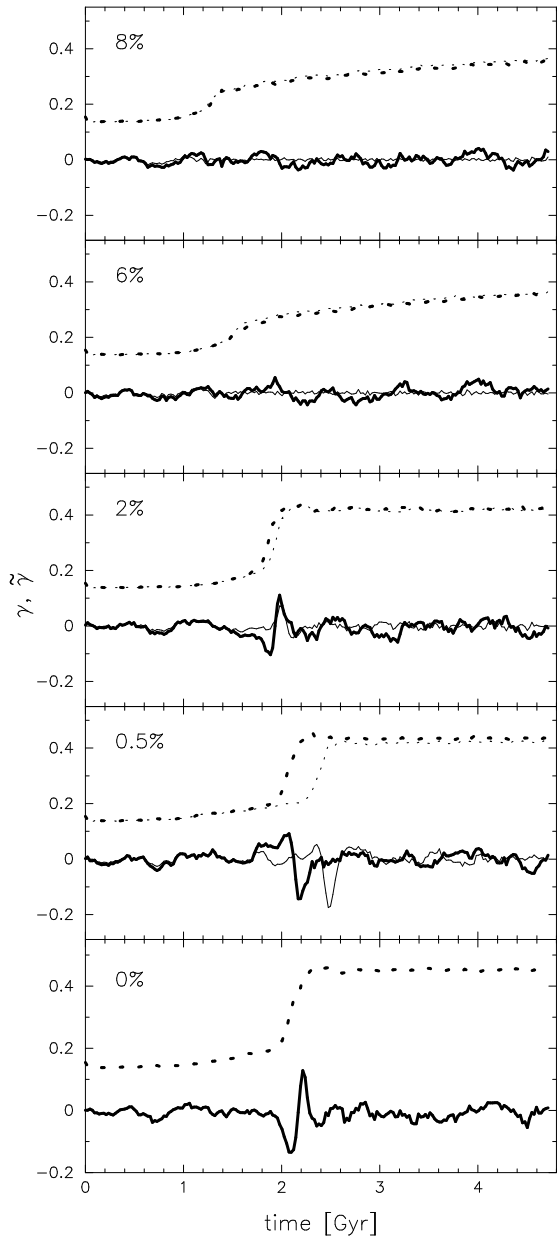}
\end{center}
\caption{Vertical thickening of the bar $\gamma\equiv (1/N_0)\Sigma_{\rm i} 
z_{\rm i}/r_{\rm i}$ (solid lines) and $\tilde{\gamma}\equiv (1/N_0)\Sigma |z_i/r_i|$ 
(dashed lines) in models with $f_{\rm gas}$ of 8\% to 0\% (from top to bottom). Thin 
and thick lines correspond to models with and without the SBH.
}
\end{figure}

Additional models testing the importance of various
parameters for the dynamical and secular evolution of bars have been advanced. 
First, $\sim 25\%-100\%$ 
of the gravitational torques, that the gas exerts on stars, have been removed
in our most gas-rich model with $f_{\rm gas}=8\%$. This barely affected the $A_2$
curve. Second, we have added the gas `force field' from our gas-rich model to the
purely stellar model --- the $A_2$ curve switched its behavior and closely
followed the original gas model preserving the same drop in $A_2$. Third,
we run a number of pure stellar models, which have grown the SBHs and the CMCs 
artificially, using their histories from the gas-rich model. No gravitational
torques from the gas onto the stars have been present, but the bar strength
has shown the same sharp decay. Fourth, pure stellar 
and gas-rich models have been run in hotter disks with initial $Q=1.8$ --- those 
have shown qualitatively similar behavior, with bars reaching
lower amplitudes  but exhibiting the same drop in $A_2$. Finally, the amount of
gas $J$ transferred to the stellar bar is small compared to the overall $J$ balance
there. 

The conclusion which emerges from these runs is that a {\it direct} input of the 
angular momentum from the gas into the stellar bar is not responsible for the sharp
drop in its strength, even in the gas-rich models. It is most revealing that the 
pure stellar model shows no 
qualitatively different behavior from models with various gas fractions. Only with 
an additional analysis of the bar structure, the differences in the evolution 
between the gas-pure and gas-rich models start to emerge, as we discuss below.

\subsection{Pure stellar models: drop in the bar strength}

We now attempt to address the issue of what is the origin of the sharp weakening
of the bar in G05\,--\,G8 and G05BH\,--\,G8BH models with gas. The pure stellar
model G0 acts here as the Rosette stone --- the reason for the drop in its 
amplitude, $A_2$, is the increasing fraction of chaotic orbits within the bar 
(Martinez-Valpuesta \& Shlosman 2004). This behavior is triggered
by the dynamical (buckling) instability, first detected 
in Combes \& Sanders (1981), and analyzed by Combes et al. (1990), Pfenniger \&
Friedli (1991), Raha et al. (1991) and others, largely based 
on Toomre (1966) interpretation. The buckling instability is 
characterized by a spontaneous break of the symmetry with respect to the disk 
equatorial plane. It results in the vertical thickening of 
the bar which acquires a characteristic boxy/peanut shape, frequently observed in 
edge-on disks along the bar minor axis (e.g., L\"utticke, Dettmar \& Pohlen 
2000).

However, formation of the characteristic boxy/peanut shape of the inner bar does not 
necessitate
the buckling instability. Friedli \& Pfenniger (1990) have shown that the near suppression 
of the vertical asymmetry in the bar still results in this characteristic shape, 
albeit established on a much longer {\it secular} timescale. They found that the bar 
thickening can be a direct consequence of the heating in the bar by its vertical 
inner Lindblad resonance (ILR) 
and other lower resonances that scatter the stellar particles out of the disk plane. 

These two alternatives for the bar to obtain the characteristic peanut shape can be
reconciled. The initial growth of the stellar bar is accompanied by a strong increase 
in the chaotic motions in the $xy$-plane, especially in its outer part beyond the 
vertical ILR and close 
to CR. These motions are the prime reason for the outer bar dissolution and 
its {\it overall} weakening in the $xy$ plane --- the bar 
actually shortens dramatically over a fraction of 
a Gyr.  What is most important is that {\it the role of the buckling is to accelerate 
the otherwise secular (vertical) heating to a dynamical timescale} (Martinez-Valpuesta 
\& Shlosman 2004). 
Particles which normally confined to the bar equatorial plane are injected 
above it, allowing them to explore a larger configuration space.    
Hence one can distinguish between the buckling which is a dynamical 
instability and the action of the resonances which drive a slow vertical diffusion of 
the planar stellar orbits. 

The model G0 exhibits just such a behavior that weakens the bar in the $xy$ plane but
does not lead to a complete bar dissolution. It displays a strong vertical asymmetry 
at $\tau\sim 2-2.3$~Gyr as given by the (vertical) Fourier coefficient $A_{\rm z,1}$ 
which has a maximum at this time (Fig.~10a) and the vertical bloating given
by another Fourier coefficient in the vertical plane, $A_{\rm z,2}$ (Fig.~10b). The 
initial growth of the 
stellar bar triggers two processes which appear to be nearly fatal for the bar itself. 
First, a larger fraction of stellar orbits in the bar becomes chaotic. The 
readily developing vertical ILR is located within the bar (typically half-way to the
CR radius) and efficiently scatters (randomizes) the orbits in the bar 
vertical plane. Particles whose energy allows them to visit the outer 
part of the bar inevitably will cross the resonance region and will be affected most, 
thus dissolving the outer half of the bar. It is possible, but remains to be proven,
that the first particles scattered above the disk plane act as `seeds' for the
(collective) buckling instability. This would explain why the appearance of the
vertical ILR and the onset of the dynamic instability happen so close in time.

Strengthening of a stellar bar increases the importance of the vertical ILR and other 
lower resonances --- this widens the associate resonance gaps in the characteristic 
diagrams.\footnote{These diagrams plot the $y$-intersection of a stellar orbit against 
the associated integral of motion, the Jacobi energy, $E_{\rm J}$ (e.g., Binney \& 
Tremaine 1987).} The ILR gap gives rise to specific symmetric/antisymmetric orbital 
families, called BAN/ABAN\footnote{These are $2:2:1$ orbits, i.e., two radial 
oscillations for two vertical oscillations for one azimuthal turn.} correspondingly 
(Pfenniger \& Friedli 1991; Berentzen et al. 1998; Skokos et al. 2002; 
Martinez-Valpuesta et al. 
2006). These orbits are the 3-D counterparts of the planar $x_1$ orbits introduced by 
Contopoulos \& Papayannopoulos (1980) and become populated when the planar orbits are 
destabilized. The BAN/ABAN orbits imprint their characteristic boxy/peanut shapes on the 
(inner) bar, in tandem with other 3-D families that originate at lower resonances. 
When the vertical symmetry in the bar is enforced, 
the population  growth on these orbits is a slow secular process developing over many 
rotations of the bar. 
 
\subsection{Gas models: drop in the bar strength and quenching the bar buckling}

All models with the gas, shown in sections~3 and 4, exhibit a drop in $A_2$ that is 
similar to the drop in the pure stellar model. Because the initial conditions for all 
these models are the same, and the evolution towards the peak in $A_2$  
and the subsequent drop are nearly identical, it is tempting to assume that the
same population of stellar orbits leaves the bar either as a result of the buckling
or the action of the CMC. This population has been identified by Martinez-Valpuesta
\& Shlosman (2004) as consisting mainly of chaotic orbits developing in
the bar mid-plane. We do not pursue this line further here. The orbital nomenclatures
are discussed in Berentzen et al. (1998) and in Patsis et al. (2002).

Evolution of the bar vertical shape is further complicated by the presence of
the gas component. The amplitude of the
vertical buckling, as measured by $A_{\rm z,1}$, shows a two-fold behavior: 
gas-poor models exhibit a substantial buckling, while gas-rich models remain
nearly symmetric (Fig.~10a). Berentzen et al. (1998)
found that, in the presence of gas, the vertical instability in the bar is damped
substantially. The new detail which emerges here is that, {\it while the bar remains 
symmetric in the gas-rich models, it nevertheless thickens}. In the gas-poor 
models, the bars thicken abruptly, and subsequently remain
unchanged, in the sense that the vertical swelling saturates immediately thereafter. 
In the gas-rich models, one can clearly distinguish two 
phases  --- an initial and fast swelling and a subsequent increase in the vertical 
thickness. This bimodal behavior is clearly displayed in Figs.~10b and 11.
The former figure contrasts the vertical $A_{\rm z,2}$ coefficient in G0 and gas
models --- the bulge forms nevertheless, but its boxy/peanut shape becomes progressively
less prominent (as observed and quantified first by Berentzen et al. 1998, and confirmed 
by Athanassoula et al. 2005). Fig.~11 supports this conclusion through 
the measure of a new parameter 
$\gamma \equiv (1/N_0) \sum_i^{} z_{\rm i}/r_{\rm i}$ and its counterpart 
$\tilde{\gamma}\equiv (1/N_0) \sum_i^{} |z_{\rm i}/r_{\rm i}|$ 
for the stellar particles in 
the more gas-rich models, with and without the BHs.  Here we sum over the 
cylindrical region of $0\leq r\leq 10$~kpc and $|z|\leq 1$~kpc, and $N_0$ is the 
number of stellar particles within this region used here for a normalization. These 
parameters quantify the 
(vertically) asymmetric and symmetric particle distributions respectively. They
appear less noisy than $A_{\rm z,1}$.

The trend  in $A_{\rm z,1}$, $A_{\rm z,2}$, $\gamma$ and $\tilde{\gamma}$ that 
separates the gas-rich from the 
gas-poor models can be explained by the presence of the CMC. The growth of the CMC
in our models is clearly linked to the gas, as the purely stellar model forms
virtually no CMC.\footnote{The only contribution to the CMC in the stellar models comes
from the stellar and DM ghost bars.} On the other hand, the final mass of the CMC 
depends linearly 
on $f_{\rm gas}$ in our models. Such massive CMCs will have immediate implications 
on the formation and shapes of galactic bulges and on the onset of the bar buckling. 
They destabilize 
the BAN/ABAN orbits, whose stable regions shrink towards the disk mid-plane ---
this damps the vertical asymmetry, reducing support for the boxy/peanut bulge
shapes (Berentzen et al. 1998; Athanassoula et al. 2005).
The increase in the stellar velocity dispersions
proceeds both on dynamical and secular timescales, and is driven by
resonant and non-resonant mechanisms, as we discuss below.

The action of the vertical resonances follows from the spatial part of the 
stellar distribution function, while its kinematic part drives the stellar 
vertical-to-radial velocity dispersion ratio, 
$\beta\equiv \sigma_{\rm z}^2/\sigma_{\rm r}^2$. The latter behavior is displayed 
in Fig.~12. We find that $\sigma_{\rm r}$ initially increases with time, 
which is related to the bar strengthening. At $\tau\sim 2-2.3$~Gyr, for G0 and
the gas-poor models, $\beta$ increases sharply from $\sim 0.3$ to $\sim 0.9$. This 
jump brings $\beta$ to nearly isotropic conditions, and its mere appearance 
signifies a sudden change in the vertical structure of the bar --- stellar settling 
on the BAN/ABAN orbits. The subsequent gradual decrease in $\beta$ follows from a 
secular increase in $\sigma_{\rm r}$ and is related to the slow growth of the bar.

While an exact value of $\beta$, when the buckling instability is triggered,
can depend on a particular model, within limits, we find that it lies around
$\beta\sim 0.4$ for models presented here and in Martinez-Valpuesta et al. (2006).
This value agrees well with that of Sellwood (1996) who found that some of his
models remained unstable for up to $\beta\sim 0.4$. 

The gas-rich models behave systematically different from the gas-poor models in 
Fig.~12, as much as they differ in Figs.~10 and 11. While $\beta$ still
drops below the threshold of 0.4, as the buckling develops, a steep growth 
of the CMCs in these models produces an equally abrupt heating of the stellar `fluid' 
--- an increase in the stellar dispersion velocities within the central kpc.
This drives  $\beta$ above the threshold and suppresses the buckling instability ---
a non-resonant effect. The comparative increase in $\sigma_{\rm r}$ prevents
$\beta$ from growing to values attained in the gas-poor models.
It is of a prime importance, that the vertical heating 
of $\sigma_{\rm z}$ is independent of $f_{\rm gas}$, in tandem with the similar 
behavior of $A_2$ in all models. The subsequent secular increase in $\beta$ is 
driven by an increase in $\sigma_{\rm z}$ and is related to the action of the 
vertical resonances discussed earlier. 

In a number of control runs (e.g., section~4), we have added the CMCs of various 
mass and compactness to the pure stellar model. The resulting $A_{\rm z,1}$ 
and $A_{\rm z,2}$ 
evolution remained two-fold and followed either the gas-poor or gas-rich models,
although no gravitational torques from the gas onto the stars were involved.
Models G0BH and G8BH250 behave as gas-poor, while G0BH80 and G0BH750 as gas-rich,
displaying substantial or negligible vertical asymmetries correspondingly, but the 
overall thickening of the bar at the end of the simulations is similar to that in 
the pure stellar model. 

\subsection{The bulge shape: effect of the gas}

The gradual loss of the characteristic shape of the bulge is 
correlated with increasing $f_{\rm gas}$ in our models. The peanut shape
is replaced by the boxy one and by an increasingly elliptical bulge (especially
in the inner isodensities). In the gas-poor models, the increase in the
velocity dispersion discussed earlier injects stellar particles above the disk.
During the buckling instability, a spontaneous breakup of the vertical 
symmetry occurs, $\sigma_{\rm z}$ increases and $\sigma_{\rm r}$ decreases
(because the hottest particles in the $xy$ plane leave it). The vertical velocity 
dispersion (i.e., degree of freedom), $\sigma_{\rm z}$, in the disk is an adiabatic 
invariant and is affected by this dynamical instability.

For the gas-rich models, the steep increase in the CMC mass (see Figs.~2 and 3) 
pumps energy both in the vertical and horizontal planes --- both $\sigma_{\rm z}$ 
and $\sigma_{\rm r}$ increase simultaneously. This heating of the stellar disk 
in the $xy$ and vertical planes is the result of a strong {\it planar} ILR 
which develops in the gas-rich models in addition to the vertical 
ILR that is present even in the gas-poor models. This was explicitly shown by
Berentzen et al. (1998) for both types of models using identical initial conditions
to those presented here. As a result, the Jacobi energy of stellar particles --- 
normally a conserved quantity, increases abruptly, and particles are injected on 
3-D orbits above the disk which are not trapped by the BAN/ABAN family. These 
orbits reduce support for the boxy/peanut-shape of the bulge which becomes 
progressively rounder with $f_{\rm gas}$, as observed in our simulations. 

\begin{figure}
\begin{center}
\includegraphics[angle=-90,width=0.95\columnwidth]{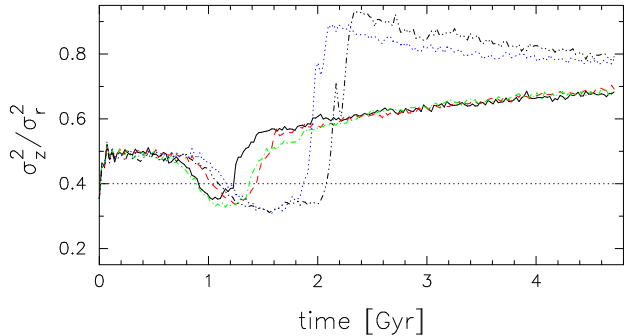}
\end{center}
\caption{Evolution of $\beta\equiv \sigma_{\rm z}^2/\sigma_{\rm r}^2$ --- the ratio 
of vertical dispersion velocities in G8BH (solid black), G6BH (dashed red), G4BH 
(dash-dotted green), G2BH (dotted blue) and G0 (black dash-dot-dot-dotted) models. 
$\beta$ is calculated
using dispersion velocities within the central kpc. The dotted line at $\beta=0.4$
corresponds to the buckling instability limit for a wide range of models here and 
elsewhere.}
\end{figure}

An additional effect can contribute further to the damping of the vertical
asymmetry in the gas-rich models. The condition for a linear vertical buckling of thin 
sheets with radial dispersion velocities has been analyzed by Toomre (1966). For pure 
collisionless systems, the instability saturates in the non-linear regime because of 
wave-particle and wave-wave damping (e.g., Sellwood, Nelson \& Tremaine 1998). In the 
presence of a substantial gas component, a strong damping originates in the two-fluid
gas-stars, interaction. The source of this damping is in the overall response
mismatch between the collisionless disk and the viscous gas. The difference in the
stellar and gaseous responses can be noticed already in the linear regime, by
applying the epicyclic approximation. Using Heller \& Shlosman (1996) notation, we 
follow the vertical oscillations at some radius $r$, irregardless of the excitation 
cause. While the stellar component follows the
harmonic oscillator equation ${\ddot Z}_1(\tau) + \nu^2 Z_1(\tau) = 0$, 
the gas follows the damped oscillator equation, ${\ddot Z}_1(\tau)+ \lambda 
{\dot Z}_1(\tau) + \nu^2 Z_1(\tau) = 0$, where the subscript `1' represents the
first order terms, $\nu$ is the vertical epicyclic frequency, $\lambda$ ($>0$) is
the damping coefficient in the gas, and the dots denote the time derivatives. 
The formal solution to the second equation is 
$Z_1(\tau)=A(\tau,\lambda) {\rm cos}(\omega\tau + \delta)$, where $A(\tau,\lambda)$ 
is the time-dependent (decaying) amplitude of the
oscillation, $\omega=\omega(\nu,\lambda)$, and $\delta$ provides the phase-shift 
between the gaseous and stellar responses. This phase-shift depends explicitly on 
the $\lambda$ and disappears when $\lambda\rightarrow 0$. 

The vertical oscillations which accompany the buckling instability are clearly
visible in the numerical simulations of a pure stellar disk embedded in a live
DM halo of Martinez-Valpuesta et al. (2006, see the accompanying animation) and
we observe them here as well. The gas component buckles as well but immediately
falls back to the disk mid-plane and remains unperturbed thereafter.
The reason for this difference in behavior is the dissipation present in the gas
--- this prevents the BAN/ABAN orbits, that have sharp turns, from sustaining 
the gas. 

Hence, the action of the CMC is to suppress 
the vertical buckling in stellar bars with larger gas fractions, above few percent 
of the disk mass, and to increase the stellar dispersion velocities. This 
weakens the boxy/peanut shape of the bulge and shortens the plateau in $A_2$. In 
the long run,
the inner bar thickens additionally due to the secular action of the vertical
resonances. So the development of the boxy/peanut bulges appears to be limited by 
the gas influx to the center in the gas-rich disks. This has been noted by
Berentzen et al. (1998) using a lower resolution model, and by Athanassoula et
al. (2005) using an analytical potential for the CMC. In the latter work, the
CMC has been introduced after the boxy/peanut-shaped bulge has formed, while
here we find that the growth of the CMC can weaken the orbital support for
this shape in the first place. Of course the star formation
will be important in altering the gas accumulation in the center and will affect
the bulge morphology in some way. Future numerical modeling will quantify this 
process. 

This means that the galactic stellar bars go through vigorous evolution not only
in the equatorial disk plane but also in the plane normal to the disk. Geometrically
thin bars in {\it both} planes readily develop a large population of chaotic orbits
which `leak' and thicken the bar over (at least) secular  but sometime dynamical
timescales. The gas inflow towards the central few hundred pc, while damping the
vertical buckling, heats up the stellar disk --- this is a vigorous heating
which proceeds on dynamical and secular timescales. We find that $\beta$ still drops 
below 0.4 (Fig.~12), and that the vertical heating is not driven by the spiral 
structure in the disk as claimed by Debattista et al. (2006) --- the heating exists 
even in a pure stellar model with a massive CMC which has no spiral structure. 
Furthermore, while the gas of course participates in the buckling itself, 
the gas layer remains thin and quickly collapses to the disk mid-plane --- this 
result is not related to the isothermal EOS and persist for adiabatic 
gas as well. The only residual swelling of the gas layer comes from its response 
to the background stellar potential which becomes shallower with time. 

The effect of the CMCs on the evolution of stellar bars in our simulations is important
in determining the bar vertical and planar structure over the simulation time.
Although we have limited the length of the evolution to $\sim 5$~Gyr, this is a substantial
period of time to assess the immediate influence of the CMCs. Beyond shortening the
plateau around the maximum in $A_2$, the main difference between
gas-poor and gas-rich models lies in that former show bars which resume their growth
after the period of buckling instability. The latter models display bars which exhibit
a mild weakening. 

The SBH masses in our simulations attain $\sim 0.3\%-5\%$ of the disk mass, depending
on $f_{\rm gas}$. From the observational point of view, this appears to be a factor 
of $\sim 10$ in excess of the SBH masses
observed in disk galaxies. What is interesting is that even these 
large masses do not lead to a dissolution of stellar bars even over the time periods
of 5~Gyr. This result is in sharp contrast with modeled bars of Friedli (1994) which decay 
completely over $1-2$~Gyr, but in agreement with Athanassoula et al. (2005) which 
report much more robust bars. One possible explanation lies in the absence of a DM 
halo in the former simulations. Under these conditions, the stellar bars develop faster,
become stronger and have a larger fraction of chaotic orbits. The dissolving action of 
the SBH will be much more formidable in this case.
  
\subsection{Summary}

In summary, galactic bars in our simulations go through various stages of evolution,
and we focus primarily on changes in the bar mid- and vertical planes. The pure 
stellar bars have been analyzed in this context in the literature (e.g., Chapter~1) 
--- our goal was to understand how the gas presence modifies this evolution, in the 
range of $f_{\rm gas}\ltorder 8\%$. 
A large number of models of a two-component disk embedded in the live DM halo 
has been analyzed. We find a two-fold evolution and contrast the gas-poor, 
$f_{\rm gas} < 3\%$ with the gas-rich, $f_{\rm gas} > 3\%$ models. The exact
dividing line, $f_{\rm gas}$, between these groups can vary but the essence remains.

First, the angular momentum transfer from the gas to the stellar bar has no
visible effect on the evolution of the bar strength in our models, beyond a 
well-known buildup of 
the CMC and the SBH there, contrary to Bournaud et al. (2005). We find that more massive
CMCs shorten dramatically the extent of the plateau near the maxima of the bar strength.
Second, all stellar bars thicken vertically, but the reason is two-fold --- gas-poor
models buckle while gas-rich models swell by preserving their symmetry. 
The vertical asymmetry (buckling) of the bar is damped in gas-rich models 
due to the forming CMC. Third, the vertical swelling starts earlier for the 
gas-rich disks and this effect increases with  $f_{\rm gas}$. Fourth, the vertical 
thickening of a stellar bar proceeds in two stages. Namely, the CMC heats up the 
central kpc in the stellar disk on a dynamical timescale, this stabilizes the bar 
against buckling, but puffs it up. This is followed by a {\it slow} (secular) stage of 
the bar thickening that complements the dynamical stage. Overall,
the degree of the stellar bar thickening is practically {\it independent} of 
the gas fraction in the disk, for $f_{\rm gas} \ltorder 8\%$.  
Fifth, the action of the CMC leads to the formation of a weaker peanut and more 
elliptical bulge, in a contrast to the boxy/peanut-shaped bulge
forming in the gas-poor models, confirming the earlier result of Berentzen et al. 
(1998). The severity of this effect depends on the star formation
process and the feedback from stellar evolution.

\acknowledgments
We acknowledge helpful conversations with Francoise Combes and Daniel Pfenniger. 
We thank Barbara Pichardo for helping with the initial stages of this project.
This research has been partially supported by NASA/LTSA 5-13063, NASA/ATP NAG5-10823, 
HST/AR-10284 (to IS), and by NSF/AST 02-06251 (to CH and IS). IB
acknowledges financial support by the project {\sc 'GRACE'} I/80\,041 
of the Volkswagen Foundation. IM-V is grateful for support by the Gruber Foundation.


\begin{thebibliography}{}

\bibitem[]{}Athanassoula, E., Sellwood, J.A. 1983, Internal Kinematics and Dynamics 
     of Galaxies, (Dordrecht: D. Reidel Publishing Co.), p.~203

\bibitem[]{}Athanassoula, E. 1992, \mnras, 259, 345

\bibitem[]{}Athanassoula, E., Misiriotis, A. 2002, MNRAS, 330, 35 

\bibitem[Athanassoula(2003)]{Ath03}Athanassoula, E. 2003, \mnras, 341, 1179 

\bibitem[]{}Athanassoula, E. 2005, in nonlinear Dynamics in Astronomy 
     \& Physics, ed. S.T. Gottesman, J.R. Buchler \& M.E. Mahon, Ann. NY 
     Acad. Sci., 1045, 168
     
\bibitem[]{}Athanassoula, E., Lambert, J.C., Dehnen, W.	2005, \mnras, 363, 496     

\bibitem[]{}Barnes, J., Hut, P. 1986, Nature, 324, 446

\bibitem[]{}Begelman, M.C., Volonteri, M., Rees, M.J.  2006, \mnras, 370, 289   
     
\bibitem[]{}Berentzen, I., Heller, C.H., Shlosman, I., Fricke, K. 1998, MNRAS, 
     300, 49 

\bibitem[]{}Berentzen, I., Athanassoula, E., Heller, C.H., Fricke, K.J. 2004,
     MNRAS, 347, 220     
     
\bibitem[]{}Binney, J., Tremaine, S. 1987, Galactic Dynamics, Princeton U. Press
     
\bibitem[]{}Bournaud, F., Combes, F., Sememlin, B. 2005, MNRAS, 364, L18 


\bibitem[]{}Buta, R., Combes, F. 1996, Fund. Cosmic Phys., 17, 95

\bibitem[]{}Combes, F., Sanders, R.H. 1981, A\&A, 96, 164

\bibitem[]{}Combes, F., Debbasch, F., Friedli, D., Pfenniger, D. 1990, 
     A\&A, 233, 82

\bibitem[]{}Contopoulos, G., Papayannopoulos, T. 1980, A\&A, 92, 33      
     
\bibitem[]{}Debattista, V.C., Mayer, L., Carollo, C.M., Moore, B., Wadsley, J.,
     Quinn, T. 2006, \apj. 645, 209

\bibitem[]{}Dehnen, W. 2002, J. Comp. Phys., 179, 27

\bibitem[]{}Englmaier, P., Shlosman, I. 2000, \apj, 528, 677

\bibitem[]{}Fall, S. M., Efstathiou, G. 1980, MNRAS, 193, 189

\bibitem[]{}Ferrarese, L., Ford, H. 2005, Space Sci. Rev., 116, 523


\bibitem[]{}Forbes, D.A., Norris, R.P., Williger, G.M., Smith, R.C. 1994,
     AJ, 107, 984

\bibitem[]{}Friedli, D., Pfenniger, D. 1990, in ESO/CTIO Workshop on Bulges of 
     Galaxies, ed. B. Jarvis \& D.M. Terndrup (Garching: ESO), 265 

 \bibitem[]{}Friedli, D., Martinet, L. 1993, A\&A, 277, 27       
     
\bibitem[]{}Friedli, D. 1994, in Mass-Transfer Induced Activity in Galaxies,
     ed. I. Shlosman (CUP), 268  
     
\bibitem[]{}Hasan, H., Norman, C.A. 1990, \apj, 361, 69      
     
\bibitem[]{}Heller, C.H., Shlosman, I. 1994, \apj, 424, 84

\bibitem[]{}Heller, C.H. Shlosman, I. 1996, \apj, 471, 143

\bibitem[]{}Heller, C.H., Shlosman, I., Englmaier, P. 2001, \apj, 553, 661.

\bibitem[]{}Heller, C.H., Shlosman, I., Athanassoula, E. 2007, ApJ Lett., 657, L65

\bibitem[]{}Ishizuki, S. Kawabe, R., Ishiguro, M., Okumura, K.S., Kasuga, T.,
    Chikada, Y., Takashi, K. 1990, Nature, 334, 224

\bibitem[]{}Jogee, S., Shlosman, I., Laine, S., Englmaier, P., Knapen, J.H., 
    Scoville, N.Z., Wilson, C.D. 2002, \apj, 575, 156 
    
\bibitem[]{}Jogee, S. 2006, Lect. Notes Phys. 693, 143    
    
\bibitem[]{}Kenney, J.D.P., Wison, C.D., Scoville, N.Z., Devereux, N.A., Young, J.S.
    1992, \apj, 395, L79    

\bibitem[]{} Knapen, J.H., Beckman, J.E., Heller, C.H., Shlosman, I., de Jong, 
    R.S. 1995, \apj, 454, 623

\bibitem[]{}Knapen, J.H. 2005, A\&A, 429, 141

\bibitem[]{}Kormendy, J., Kennicutt, R.C. 2004, ARA\&A, 42, 603

\bibitem[]{}L\"utticke, R., Dettmar, R.-J., Pohlen, M. 2000, A\&A, 145, 405

\bibitem[]{}Lynden-Bell, D., Kalnajs, A.J. 1972, MNRAS, 157, L1

\bibitem[]{}Maiolino, R., Alonso-Herrero, A., Anders, S., Quillen, A., Rieke, M.J.,
     Rieke, G.H., Tacconi-Garman, L.E. 2000, \apj, 531, 219 

\bibitem[]{}Martinez-Valpuesta, I., Shlosman, I. 2004, \apj, 637, 214

\bibitem[]{}Martinez-Valpuesta, I., Shlosman, I. Heller, C.H. 2006, \apj, 613, L29

\bibitem[]{}Norman, C.A., Sellwood, J.A., Hasan, H. 1996, \apj, 462, 114

\bibitem[]{}Ostriker, J.P., Peebles, P.J.E. 1973, \apj, 186, 467

\bibitem[]{}Patsis, P.A., Skokos, Ch., Athanassoula, E. 2002, MNRAS, 337, 578

\bibitem[]{}Pfenniger, D., Friedli, D. 1991, A\&A, 252, 75

\bibitem[]{}Raha, N., Sellwood, J., James, R.A., Kahn, F.D. 1991, Nature, 352, 411

\bibitem[]{}Sellwood, J.A. 1996, in Barred Galaxies, eds. R. Buta, D.A. Crocker
    \& B.G. Elmegreen (San-Francisco: ASP), 259 

\bibitem[]{}Sellwood, J.A., Nelson, R.W., Tremaine, S. 1998, \apj, 506, 590

\bibitem[]{}Sellwood, J.A. 2006, \apj, 637, 567

\bibitem[]{}Shen, J., Sellwood, J.A. 2004, \apj, 604, 614

\bibitem[]{}Shlosman, I., Frank, J., Begelman, M.C. 1989, Nature, 338, 45

\bibitem[]{}Shlosman, I., Begelman, M.C., Frank, J. 1990, Nature, 345, 679

\bibitem[]{}Shlosman, I., Noguchi, M. 1993, \apj, 414, 474

\bibitem[]{}Shlosman, I., Heller, C.H. 2002, \apj, 565, 921
 
\bibitem[]{}Shlosman, I. 2005, The Evolution of Starbursts, S. Huettemeister \& 
    E. Manthey (eds.), Melville: AIP, 223

\bibitem[]{}Skokos, Ch., Patsis, P.A., Athanassoula, E. 2002, MNRAS, 333, 847    
    
\bibitem[]{}Toomre, A. 1966, in Notes on the Summer School Program in Geophysical
    Fluid Dynamics at the Woods Hole Oceanographic Institution, Vol.~66, 111

\bibitem[]{}Tremaine, S., Weinberg, M.D. 1984, MNRAS, 209, 729

\bibitem[]{}Weinberg, M.D. 1985, MNRAS, 213, 451

\end{thebibliography}
\end{document}